# Rule- and dictionary-based solution for variations in written Arabic names in social networks, big data, accounting systems, and large databases

Ahmad B. A. Hassanat[1], Ghada Awad Altarawneh[2]
IT Department[1], Accounting Department[2],
Mu'tah University, Mu'tah – Karak, Jordan, 61710
ahmad.hassanat@gmail.com[1], ghadatrn@yahoo.com[2]

**Abstract**

This paper investigates the problem that some Arabic names can be written in multiple ways. When someone searches for only one form of a name, neither exact nor approximate matching is appropriate for returning the multiple variants of the name. Exact matching requires the user to enter all forms of the name for the search, and approximate matching yields names not among the variations of the one being sought.

In this paper, we attempt to solve the problem with a dictionary of all Arabic names mapped to their different (alternative) writing forms. We generated alternatives based on rules we derived from reviewing the first names of 9.9 million citizens and former citizens of Jordan. This dictionary can be used for both standardizing the written form when inserting a new name into a database and for searching for the name and all its alternative written forms.

Creating the dictionary automatically based on rules resulted in at least 7% erroneous acceptance errors and 7.9% erroneous rejection errors. We addressed the errors by manually editing the dictionary. The dictionary can be of help to real world-databases, with the qualification that manual editing does not guarantee 100% correctness.

## 1. Introduction

Online social networks, accounting systems, and large database systems allow people to register their names any preprocessing being done to the names, except for the imposition of a few rules, such as the exclusion of numbers and special characters from the names.

When the names being entered in Arabic and the people entering them come from different backgrounds, they enter the names with their local accents and understanding, and after a short period of time, the same name is entered in different forms. For example, the Arabic name "*Rola*" "رولا" might also be entered using characters such as "رلی"//"Rla," "روله"//"Rulah," "رلا"//"Rlaa," "رُلا"//"Rulaa," "رُلی"//"Rula," and so on. A Google search for these six variants resulted in 4,420,000, 356,000, 91,400, 218,000, 20,800, and 5,560 hits, respectively.

Without having a solution to this problem, when someone searches for such a name, that person must use exactly the variant written and stored in the database. If a banker is asked to search for the name of a customer and the banker enters the wrong alternative form, the result might be embarrassing to the banker and dissatisfactory to the customer, if the customer cannot remember the account number. The problem is more acute in accounting systems, when use of the wrong variant may result in payments being denied if, for example, the name written on a check or a bill is in a different form than the one on the ID.

Another example—if someone searches for a friend on Facebook® and knows the person's name, but not the exact variant, the searcher must try all alternatives manually. Even if the searcher knows all of the variants, the process may be time-consuming.

Diversity in the writing of Arabic names is due to several factors, such as ambiguity in Arabic morphology, ambiguity in Arabic orthography, and insufficient and ambiguous methods of standardizing Arabic names. To solve this problem, this paper presents a rule-based and dictionary-based solution for writing and searching for Arabic names. We created the dictionary using all of the Arabic first names generously provided to us by the Jordanian civil status and passport department. The names include 75,000 distinct row names belonging to 9.9 million (living and dead) citizens.

Edit distance is employed (offline) to compare each name with the others to collect all similar names—those with distances less than or equal to one. Then we confirmed the similarity using some rules related to the reasons behind having different written forms of Arabic names[*], the confirmation continues manually to edit the resultant dictionary. Names in the resultant dictionary are indexed alphabetically to decrease searching time for names. Using edit distance alone (online) as a solution for this problem is not practical for two reasons. 1) It consumes time. 2) It yields unrelated names.

When someone searches for an Arabic name in an accounting database for example, alternative forms of the name are first sought in the proposed dictionary, and then the algorithm searches for the both the entered name and the alternatives. In addition, this dictionary is important for document matching, plagiarism detection, and recognition of Arabic names for natural language processing (NLP). The problem of matching different writing forms of Arabic names has not been the subject of sufficient studies. This paper attempts to fill that gap.

The rest of the paper is organized as follows. Next, we discuss the motivation of the study and the previous related work. Section 3 explains the proposed solution and how to use it to solve the problem in both reading and writing modes. Section 4 presents the results of applying the rules to the collected names to form the dictionary in addition to identifying and discussing more reasons for the problem. We end this paper with the main conclusions and ideas for future work.

## 2. Motivation and Previous Work

Technology related to Arabic Names is playing an increasingly important role in a variety of practical applications, such as Arabic NLP, named entity recognition, machine translation, cross-language information retrieval, and various security applications, such as anti-money laundering activities, terrorist watch lists, and criminal tracking systems.

Despite the importance of those applications, Arabic Names has not been the subject of sufficient studies that examine linguistic issues, such as problems resulting from the writing of Arabic names in different forms. Such problems flourish in the absence of standardization, the existence of different accents, and the nature of Arabic text in which words/names are written mostly without diacritics, which does not preserve the sound of the word/name. These problems lead to a situation in which Arabic writers write the same name in different forms. See table 1 for some examples of the problem.

---

[*] We mean by an "Arabic name" any name written in Arabic text, this includes foreign names transliterated to Arabic.

Table 1, examples of discrepancies in the writing of some Arabic names

| Form1 | Form2 | Form3 | Form4 |
|---|---|---|---|
| اباء | ايباء | ابا | ايبا |
| اثريا | اثريه | ثريا | ثريه |
| احميدى | احميده | احميدي | حميدى |
| خضيره | اخضيرا | اخضيره | خضيرا |
| اروى | اروه | ارواء | اروا |
| زهيا | ازهيه | ازهيا | زهيه |
| اسما | اسمه | اسمى | اسماء |
| اضحيا | اضحيه | ضحيا | ضحيه |
| اعليا | عليا | اعليه | عليه |
| يرين | ايرين | ايرينا | ايريني |
| بثينا | بثينه | بوثينه | بوثينا |
| تامارا | تمارا | تمارى | تماره |
| جمانا | جومانا | جمانه | جومانه |
| اسامه | اساما | اوسامه | اوساما |
| بهجت | بهجات | بهجه | بهجة |
| ضيف الله | ظيف الله | ضيفالله | ضيف الله |
| داود | داوود | داؤد | داؤود |
| زكريا | زكريه | زاكري | زكري |
| حسين | احسين | الحسين | حسيين |
| يحي | يحيا | يحيى | يحيي |

Arabic language is ambiguous, because Arabic names (like Arabic words generally) are written as a string of consonants with vowels usually omitted. In some cases diacritics are used to indicate short vowels, while the use of consonants to indicate long vowels is still ambiguous. This situation is major challenge for Arabic language-processing applications. There are two kinds of ambiguity—morphological ambiguity and orthographical ambiguity (Soudi, Bosch, & GUNTER, 2007) and (FARGHALY & SHAALAN, 2009).

The morphological ambiguity occurs because Arabic is an extremely inflected language. This is indicated by changing the vowel patterns and adding various suffixes and prefixes to words. For example, the name "زارع" /zarie/ (planter) might be represented in several word forms according to the context, such as "مزارع" /muzarie/ (farmer), "زرع" /zara'a/ (to plant), "زرع" /zari/ (the noun plant), "مزرعة" /mazrah/ (farm), etc. On the orthographic level, Arabic is also ambiguous. For example, the word "فلح" may theoretically represent many consonant-vowel permutations, such as *falah, faleh, faloh, falha,* etc. (Halpern, 2007).

Arabic names are no exception, regarding the ambiguity of Arabic; both kinds of ambiguity affect the written forms of names. Normally, Arabic readers use context to resolve this indistinctness, but for machines it is a non-trivial task. In describing the ambiguity, Aboaws

Alshamsan identified twenty different reasons why Arabic names are written differently, (Alshamsan, 2003; in Arabic). The following are some of the identified reasons:

1. Some names are acoustically similar: e.g., the character "س" is sometimes pronounced and written "ص," such as in the names "سلطان" and "صلطان".

2. Some characters are written similarly: e.g., the character "ى" is sometimes written "ي," such as in the names "يحيى" and "يحي".

3. The characters "ض" and "ظ" are often mixed: e.g., the name "ضياء" becomes "ظياء," because they are acoustically similar.

4. In place of "ج" writers sometimes use "ش": e.g., the name "اجدر" becomes "اشدر"; this substitution occurs due to differences in accents.

5. In place of "ا" writers sometimes use "ـه": e.g., the name "خضرا" becomes "خضره".

6. In place of "ـه" writers sometimes use "ا": e.g., the name "تاله" becomes "تالا".

7. In place of "ق" writers sometimes use "ج": e.g., the name "قاسم" becomes "جاسم." This substitution occurs due to differences in accents, usually the accents of Arab Gulf people.

8. In place of "ذ" writers sometimes use "د" or "ض": e.g., the name "ذهب" becomes "دهب," and the name "مذخر" becomes "مضخر." Acoustic similar

9. "Hamzah" (ء), is sometimes altered by deletion: e.g., the name "إبراهيم" becomes "براهيم." The name can be changed to "ي" (such as when "رائد" becomes "رايد") or to "و," (such as when "أحيسن" becomes "وحيسن").

10. The Arabic definite article "ال" (or "ام" in some accents) is sometimes added to the beginning of names: e.g., the name "أخضر" becomes "الأخضر".

11. The feminization Ta'a is sometimes added to the end of names: e.g., the name "عزة" becomes "عزت".

12. In place of "ة" writers sometimes use "ه": e.g., the name "غادة" becomes "غاده".

13. Long vowels "ا،و،ي" are sometimes added to represent short vowels "ًٌَُِّْ": e.g., the name "رندا" becomes "راندا," the name "رُبى" becomes "روبى" and the name "رهام" becomes "ريهام".

14. In place of "ا" writers sometimes use "ءا" and vice versa at the end of a name: e.g., the name "غيداء" becomes "غيدا".

15. In place of "ا" writers sometimes use "ى" (and vice versa) at the end of a name: e.g., the name "رولا" becomes "رولى".

This study uses a subset of Alshamsan's list of rules to identify different forms of the same name. Please refer to Alshamsan's work for more clarification and detailed explanations.

Another reason for variations in names (not mentioned in the previous reference) is to do with system restrictions—most large databases of Arabic names are old and built on old mainframes that use seven bits to represent characters. Seven bits is not enough to represent (in addition to English characters, numbers and special characters) Arabic diacritics and some characters such as "أ," "إ," "آ" and "ؤ." Therefore, people who entered the names in the databases had to cope by writing names differently, such as by entering the name "أحمد" as "احمد."

To the best of our knowledge, no solution has yet been proposed to address variations in Arabic names, this paper attempts to fill this gap.

A similar problem is the Romanization of Arabic names, in which several versions of the same Arabic name are transliterated in the Roman alphabet. A few solutions have been proposed to solve this problem. For example the work of (Arbabi, Fischthal, Cheng, & Bart, 1994) presents a hybrid algorithm designed to automate the transliteration of Arabic names in real time. They employed an artificial neural network (ANN) and a knowledge-based system to add vowels to Arabic names. Their ANN system filters out unreliable names, passing only the reliable names on to the knowledge-based system, which is designed for Romanization. This approach was developed at the IBM Federal Systems Company, and made available for a broad range of applications, such as for visa and document processing by border control personnel.

Related work was done by Mansour Alghamdi (Alghamdi, 2005). The work describes a new system for standardized Romanization of Arabic names. Alghamdi's system is composed of three major steps; the first step is diacritization of the Arabic word and removal of typing habits such as the insertion of dashes between letters. The second step is to convert Arabic graphemes into phonetic symbols. The third step is to convert sound symbols into the Roman alphabet. The work of (Al-Onaizan & Knight, 2002) also fits in this category.

Another similar problem is the recognition of Arabic names in a text. An example of work on this problem is the work done by (Elsebai, Meziane, & Belkredim, 2009) who used a set of keywords to guide their algorithm—the phrases that most often include a person's name. In addition to making use of the Buckwalter Arabic Morphological Analyzer (Buckwalter, 2004). The dictionary and solution presented in this paper can also be applied to the problem of recognizing Arabic names in text as part of NLP.

Our proposed method can also be used for the problem of matching personal names in English to the same names in Arabic writing. A solution to the matching problem is proposed by (Freeman, Condon, & Ackerman, 2006), who reported significant improvement by using the classic Levenshtein edit-distance algorithm.

The Levenshtein edit-distance algorithm (Levenshtein, 1966) has been used extensively in research on approximate-string matching. The algorithm compares two strings by counting the number of character insertions, deletions, and substitutions required to convert one string to a second. The question thus arises: why not using the edit-distance algorithm to approximate the searched name and view the most relative name?

This method (Edit-distance—in our case) will not work properly for two reasons. First, it is unclear what the threshold for matching should be. If the threshold is 1 (the smallest threshold), using edit-distance to search for the name "أحمد," for example, will yield the three totally unrelated names "احمد,""حمد" and "احمر". The correct search results should contain only "احمد" and its alternatives, such as "أحمد." Second, using edit distance for all names in a database is

time-consuming—the time complexity is $O(NM^2)$, where N is the number of names in the database, and M is the average length of names in the database.

## 3. The proposed solution

The proposed solution has two modes—writing and reading. In writing mode, inserting new names in a standard form in the database should be guaranteed, so the problem is solved for future searching. In reading mode, the proposed solution works for the current (corrupted) names in different databases, such as current accounting and banking systems, social media networks, and national passports systems. The solution should guarantee that all possible alternatives for the searched name are returned.

The proposed solution consists of two steps:

Step 1: The entered name is pre-processed. Pre-processing includes the removal of special characters, numbers, and spaces, except for compound names such as "عبد الله," for which one space between the two words is allowed and should be enforced. This step is performed in the same way for both reading and writing modes.

Step2: The use of our dictionary (contains all Arabic names and their alternative writing forms) made for the purpose of this study. Basically, the dictionary, is a table in a simple database, consists of all names and all their alternative writing forms, in addition to a field containing the standard writing form of each name. The standard name is the most frequent form of each name in the real database (the one that contains all Jordanian names), assuming that the most common written form is the most accurate one, because common linguistic errors often become acceptable (over time).

  a) In writing mode, when a name is inserted, the dictionary is invoked and returns the standard name as an alternative for the entered name, advising the person entering data (either a user or data-entry clerk) to insert the standard name if the entered name is corrupted (an invalid written form of the name). Figure1 depicts the proposed system in the writing mode.
  b) In the reading mode, and when a user searches for a specific name, the dictionary is invoked and passes all alternative written forms of that name to the search algorithm. The algorithm searches for the name and all alternatives using other information, such as the family name, the birth date, etc. the user can then choose the desired name. Figure2 depicts the proposed system in the reading mode.

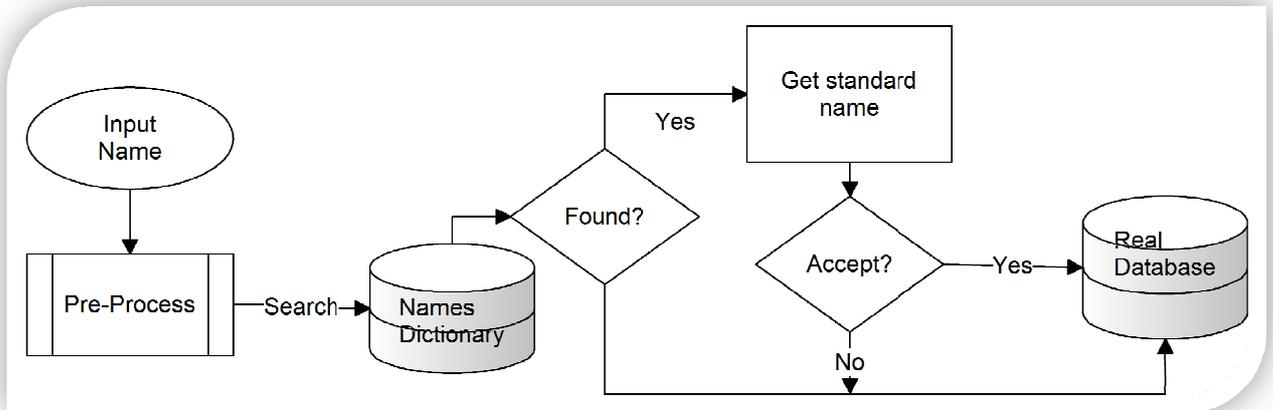

Figure 1. Flow chart of the proposed system in writing mode.

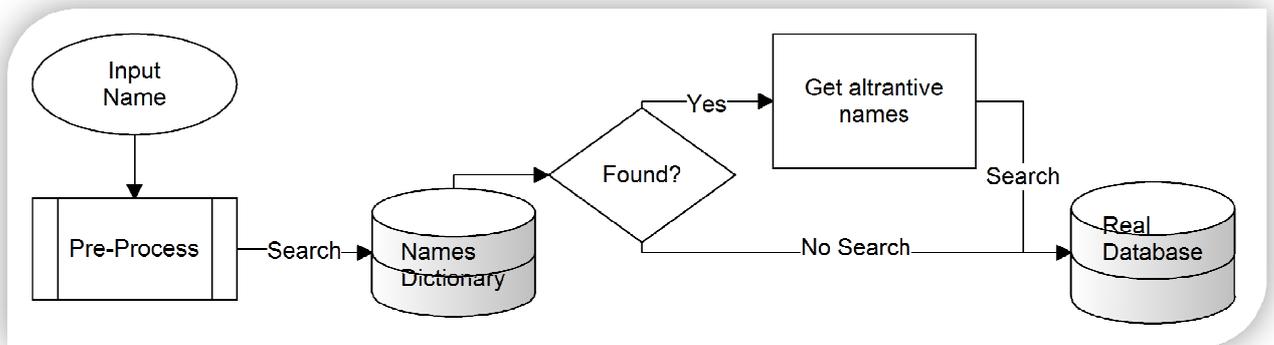

Figure 2. Flow chart of the proposed system in reading mode.

### Data collection and dictionary preparation

All Jordanian names were collected from the Jordanian civil status and passport department, who generously provided us with all the Jordanian first names in their database. The list included 75,000 distinct names; after the removal of mistakes, special characters, double middle spaces, spaces from left and right for all names, the number of distinct names was dramatically reduced to 17992. We compared those names to find the edit distance of threshold 1 based on the mentioned rules in section 2. A threshold of more than 1 allows many unrelated names to be considered as alternatives. A threshold of 2 is used only for rule 10, because this rule is a special case in which two characters ("ال") are added or removed at the beginning of a name. Alternatives come from rule 10 might be excluded from the dictionary, as this rule is easier to be checked online.

If the compared name satisfied the edit distance threshold and also one of the mentioned rules, it was then considered as an alternative to the searched name. We did this for all names in the dictionary. Algorithm 1 (written later in C++) shows the list of constraints that we used in the creation of the dictionary.

---

**Algorithm 1: create name dictionary**

---

**Input**: text file containing all Jordanian first names (17992 distinct names)
**Output**: text file represents the name dictionary which contains each name and its alternative forms
**For each** name **as** A **in** Input file **{**
    List= NULL                          //dynamic array of strings to contain the name and its alternatives.
    **Add** A to A's list
    **For each** name **as** B **in** Input file **{**
        D=EditDistance(A,B)
        **If** D=1 **{**
            **If** index(B,"س")= index(A,"ص") **or** index(A,"س")= index(B,"ص") THEN
                **ADD** B **TO** A's LIST  //Rule 1
            **If** (LastChar(B)="ي" and  LastChar (A)="ى") **or** (LastChar(B)="ى" and  LastChar (A)="ي") THEN
                **ADD** B **TO** A's LIST   //Rule 2
            **If** index(B,"ض")= index(A,"ظ") **or** index(A,"ض")= index(B,"ظ") THEN
                **ADD** B **TO** A's LIST  //Rule 3
            **If** index(B,"ج")= index(A,"ش") **or** index(A,"ج")= index(B,"ش") THEN
                **ADD** B **TO** A's LIST  //Rule 4
            **If** (LastChar(B)="ا" and  LastChar (A)="ه") **or** (LastChar(B)="ه" and  LastChar (A)="ا") THEN
                **ADD** B **TO** A's LIST  //Rule 5 & 6
            **If** index(B,"ج")= index(A,"ق") **or** index(A,"ج")= index(B,"ق") THEN
                **ADD** B **TO** A's LIST  //Rule 7
            **If** index(B,"ذ")= index(A,"د") **or** index(A,"ذ")= index(B,"د") THEN
                **ADD** B **TO** A's LIST  //Rule 8-a
            **If** index(B,"ذ")= index(A,"ض") **or** index(A,"ذ")= index(B,"ض") THEN
                **ADD** B **TO** A's LIST  //Rule 8-b
            **If** index(B,"ى")= index(A,"ي") **or** index(A,"ى")= index(B,"ي") THEN
                **ADD** B **TO** A's LIST  //Rule 9
            **If** (LastChar(B)="ة" and  LastChar (A)="ت") **or** (LastChar(B)="ت" and  LastChar (A)="ة") THEN
                **ADD** B **TO** A's LIST  //Rule 11
            **If** (LastChar(B)="ة" and  LastChar (A)="ه") **or** (LastChar(B)="ه" and  LastChar (A)="ة") THEN
                **ADD** B **TO** A's LIST  //Rule 12
            **If** (index(B,"ا") not = index(A,"ا") **and** index(B,"ا")>0) **or** (index(A,"ا") not = index(B,"ا") and index(A,"ا") >0 ) THEN
                **ADD** B **TO** A's LIST  //Rule 13-a (long vowel "ا")
            **If** (index(B,"و") not = index(A,"و") **and** index(B,"و")>0) **or** (index(A,"و") not = index(B,"و") and index(A,"و") >0 ) THEN
                **ADD** B **TO** A's LIST  //Rule 13-b (long vowel "و")
            **If** (index(B,"ي") not = index(A,"ي") **and** index(B,"ي")>0) **or** (index(A,"ي") not = index(B,"ي") and index(A,"ي") >0 ) THEN
                **ADD** B **TO** A's LIST  //Rule 13-c (long vowel "ي")
            **If** (Last2Chars(B)="اء" and  LastChar (A)="ا") **or** (LastChar(B)="ا" and  Last2Chars (A)="اء") THEN
                **ADD** B **TO** A's LIST  //Rule 14
            **If** (LastChar(B)="ا" and  LastChar (A)="ى") **or** (LastChar(B)="ى" and  LastChar (A)="ا") THEN
                **ADD** B **TO** A's LIST  //Rule 15

        **}**
        **If** D=2 **{**
            **If** (First2Chars(B)="ال" and  First2Chars (A) not ="ال") **or** (First2Chars(B) not="ال" and  First2Chars (A) ="ال") THEN
                **ADD** B **TO** A's LIST  //Rule 10
        **}**
    **}** //for B
    Write List to OUTPUT file
    Write End-line
**}** // for A
Save OUTPUT file
//end algorithm

---

## 4. Results and discussions

The dictionary that resulted from applying our algorithm to the database of names consists of one table. The table is indexed by the first field, which contains the Arabic name, and the next 13 fields contain the alternative names, with an integer field showing how frequent each alternative was in the main database. The last field contains the standard (most frequently used) form of the name.

The algorithm found 11330 names with at least one alternative. Despite the care we took in constructing rules and constraints, after manually checking the resultant dictionary, we found that some names were rejected as alternatives and others were wrongly accepted as alternatives. We therefore amended the dictionary manually to fix the erroneous acceptance and rejection errors. As a result of our manual editing, the number of names with at least one alternative rose to 11433. Names were grouped based on the number of alternatives, from 1 to 13. Table 2 shows the number of alternatives.

The number of names with 1 and 2 alternatives decreased after manual editing by 804 names. The change is due to names which satisfied the rules and Algorithm 1 constraints, but which we rejected, because they were different names. For example the name "حسن" was listed as an alternative for the different name "حسين," so we rejected it manually. By our calculation, the erroneous acceptance rate of Algorithm 1 was at least 7% for all numbers of alternatives, from 1–13.

Table 2. The number of names with each number of alternative names, before and after manual editing of the dictionary.

| Number of alternatives | Number of names before manual editing | Number of names after manual editing | Change in the number of names due to editing |
|---|---|---|---|
| 1 | 5939 | 5422 | -517 |
| 2 | 2972 | 2685 | -287 |
| 3 | 1344 | 1524 | 180 |
| 4 | 583 | 852 | 269 |
| 5 | 258 | 424 | 166 |
| 6 | 130 | 189 | 59 |
| 7 | 63 | 177 | 114 |
| 8 | 20 | 69 | 49 |
| 9 | 17 | 40 | 23 |
| 10 | 3 | 29 | 26 |
| 11 | 1 | 10 | 9 |
| 12 | 0 | 0 | 0 |
| 13 | 0 | 12 | 12 |

The number of names with 3–13 alternatives increased after manual editing by 907 names. This reflects the manual addition of alternatives that were rejected as not satisfying the rules and Algorithm 1 constraints, but are acceptable alternatives. For example, according to the rules, the female name "ادهميه" has only one alternative name: "ادهمه." However, we found and accepted two other alternatives, "ادهيمه" and "دهمه," manually. By our calculation, the erroneous rejection rate of Algorithm 1 was at least 7.9%, for all numbers of alternatives, from 1–13.

Erroneous acceptance and rejection of alternatives occurred because 1) some names satisfied the rules despite not being alternatives, 2) some acceptable alternative names did not satisfy the rules, and 3) other problems occurred, which had nothing to do with the rules, and have not been mentioned in the literature. Some of the errors of the last type include:

**1.** Spelling errors and typos, such as when the name "علاء" is written "ءلاء," and the confusion of "Hamzah" ("ء") and "Ein" ("ع"). (Many Arabic writers, including data-entry clerks, have not mastered the rules of Arabic "Hamzah".) For example, the female name "رؤى" included 13 misspelled variants: "روى", "رواى", "روا", "روئ", "روئا", "روئة", "روئه", "رؤه", "رؤا", "رؤوى", "رؤى", "روىء" and "رئى." This name is the one most often written differently. Table 3 shows other examples that fit this problem.

People employed for data entry usual have not attained a high educational level. They are trained mostly in how to use the system, and it is assumed that they can write well. Due to the ambiguity in Arabic, we believe that writing correctly in Arabic is not an easy task; therefore, data-entry clerks need to be trained to write correctly using a computer keyboard, and only then trained to use the database.

2. Foreign names are written in multiple ways in Arabic, mainly because some foreign sounds are not represented in the Arabic alphabet, such as the English letter "g" in "Margret." The letter "g" is usually represented by "ج," or "غ." In the absence of standardization, the name "Margret" and other similar names might be written, for example, as "مارجاريت" or "مارغريت." In addition, "g" is also transliterated sometimes as "ق" in Arabic, so the name "Margo" becomes "مارقو," "مارغو" or "مارجو." The same applies to the letter "s," which is sometimes transliterated as "z," so that "Teresa," for example, would be written "تريزا" with "z" or as "تيريسا" with more of an "s" sound. The letter "t" and combination "th" are also often confused, so that "Elizabeth" is transliterated to many different forms in Arabic, such as "اليزابيث" and "اليزابت." Table 4 shows some examples taken from our proposed dictionary for solving these problems.

Table 3. Some example alternative Arabic names in the dictionary that occurred due to spelling errors and typos. Empty and hidden fields mean no alternatives.

| Name | Alt1 | Alt2 | Alt3 | Alt4 | count | Standard Name |
|---|---|---|---|---|---|---|
| ءالاء | علاء | | | | 4 | علاء |
| اءبد | اعبد | | | | 9 | اعبد |
| اءتدال | اعتدال | | | | 191 | اعتدال |
| اءتماد | اعتماد | | | | 33 | اعتماد |
| اءرابي | اعرابي | | | | 4 | اعرابي |
| اءطيش | اعطيش | | | | 4 | اعطيش |
| اءمر | اعمر | | | | 3 | اعمر |
| عبدالرؤوف | عبدالرؤف | عبدالرووف | عبدالروئف | عبدالروئوف | 2915 | عبدالرؤوف |
| عبدالرؤف | عبدالرووف | عبدالرؤف | عبدالروئف | عبدالرؤوف | 9 | عبدالرؤوف |
| عبدالرؤف | عبدالرؤف | عبدالرووف | عبدالروئف | عبدالروئوف | 100 | عبدالرؤوف |
| عبدالروئف | عبدالرؤف | عبدالرووف | عبدالروئوف | عبدالرؤوف | 5 | عبدالرؤوف |
| عبدالرووف | عبدالرؤف | عبدالرؤف | عبدالروئوف | عبدالرؤوف | 209 | عبدالرؤوف |

Table 4. Examples (from our dictionary) that show alternatives of transliterated names in Arabic. Empty and hidden fields mean no additional alternatives.

| Name | Alt1 | Alt2 | Alt3 | Alt4 | Alt5 | Alt6 | Alt7 | Alt8 | count | Standard Name |
|---|---|---|---|---|---|---|---|---|---|---|
| ماريا تريز | ماريا تيريزا | ماريا تيريسا | ماريا تيريزا | مارياتريزا | ماريا تيريز | ماري تريز | ماري تريز | | 3 | ماريا تريزا |
| ماريا تريزا | ماريا تيريزا | ماريا تريز | ماريا تيريزا | مارياتريزا | ماريا تيريز | ماري تريز | | | 9 | ماريا تريزا |
| ماريا تريزا | ماريا تيريزا | ماريا تريز | ماريا تيريسا | مارياتريزا | ماريا تيريز | ماري تريز | | | 3 | ماريا تريزا |
| ماريا تيريسا | ماريا تيريزا | ماريا تريز | ماريا تيريزا | مارياتريزا | ماريا تيريز | ماري تريز | | | 3 | ماريا تريزا |
| اليزابت | اليزبتا | اليزابتا | اليزبث | اليزابيت | اليزبيت | الزابيت | اليزابيتا | | 9 | اليزابيث |
| اليزابيث | الزابيت | اليزابت | اليزبث | اليزبيت | اليزابيت | اليزبتا | اليزابتا | | 109 | اليزابيث |
| اليزابيث | اليزابت | اليزبث | اليزبيت | اليزابيتا | الزابيت | اليزبتا | اليزابتا | | 305 | اليزابيث |
| اليزبث | اليزابت | اليزبث | اليزبيت | اليزابيتا | الزابيت | اليزبتا | اليزابتا | | 11 | اليزابيث |
| مارجريت | مرجريت | مارجاريتا | مارجاريت | مارغريت | مارغريتا | مارغريتا | مرغريتا | | 102 | مرغريت |
| مارجريتا | مرجريت | مارجاريت | مارغريت | مارغريتا | مارغريتا | مرغريتا | | | 21 | مرغريت |
| مرغاريتا | مارغاريتا | مارجاريت | مارجريتا | مرجريت | مارغريتا | مرغريتا | مارغريتا | | 352 | مرغريت |
| مارغريت | مارغاريتا | مارجاريت | مارجريت | مرجريتا | مارغريتا | مرغريت | مارغريتا | | 173 | مرغريت |
| مارغو | مارقو | مارجو | | | | | | | 26 | مارغو |
| مارقو | مارجو | مارغو | | | | | | | 21 | مارغو |
| مارجو | مارقو | مارغو | | | | | | | 19 | مارغو |

3. Compound names are sometimes written with spaces between names and sometimes without. Table 5 shows examples of how our dictionary solves this problem.

Table 5. Examples (from our dictionary) that show alternatives for Arabic compound names.

| Name | Alt1 | Alt2 | Alt3 | Alt4 | count | Standard Name |
|---|---|---|---|---|---|---|
| بهاء الدين | بهاءالدين | بهاالدين | بهاع الدين | بهاعالدين | 4105 | بهاء الدين |
| بهاءالدين | بهاء الدين | بهاالدين | بهاع الدين | بهاعالدين | 560 | بهاء الدين |
| بهاالدين | بهاءالدين | بهاء الدين | بهاع الدين | بهاعالدين | 3 | بهاء الدين |
| رجاالله | رجا الله | | | | 7 | رجا الله |
| رجا الله | رجاالله | | | | 35 | رجا الله |
| رجب خان | رجبخان | | | | 11 | رجب خان |
| رجبخان | رجب خان | | | | 6 | رجب خان |
| ضيف الله | ظيف الله | ضيفالله | | | 249 | ضيف الله |
| ضيفالله | ضيف الله | ظيف الله | | | 4 | ضيف الله |
| عوده الله | العوده الله | عودهالله | | | 924 | عوده الله |
| عودهالله | عوده الله | العوده الله | | | 3 | عوده الله |
| ماشاء الله | مشاء الله | ماشاءالله | ما شاء الله | مشاء الله | 8 | ماشاء الله |
| ماشاءالله | مشاء الله | مشاءالله | ما شاء الله | مشاء الله | 5 | ماشاء الله |
| عطا الله | عطاء الله | عطاالله | عطاالله | | 4827 | عطا الله |
| عطا الله | عطاء الله | عطاالله | | | 12 | عطا الله |
| عطاالله | عطاالله | عطاء الله | عطا الله | | 1284 | عطا الله |
| عطاالله | عطاالله | عطاء الله | عطا الله | | 138 | عطا الله |

Our dictionary shows that female names include alternatives than male names, because Arabic female names usually end with "ى," "ا," "اء," "ة," or "ـه." These characters are important for distinguishing female names from male names with the same roots. For example, the male name "حسين" becomes the female name by adding "ـه" to the end to convert it to "حسينه," or adding "ا" to convert it to "حسينا." The use of many female characters results in a larger number of variations in the forms of female's names. Table 6 shows some examples of male and female names.

A great deal of effort will need to be expended to standardize Arabic names, and it will require the cooperation of all Arab countries. This paper is just an attempt to fix and standardize names taken from the Jordanian civil status and passport department. Other experimenters may use the methodology in this paper on data from other Arab countries to create a complete Arabic names dictionary. Such a dictionary would be useful for international accounting systems, social network databases, and security tracking systems.

Instead of creating a dictionary, other interested parties might also apply rules directly (online) while users search. The problem with this approach is that when those rules are applied to the searched name, they generate a large number of combinations, which is time-consuming both during the generation of combinations and search for each combination. Moreover, the combinations are not necessarily names or alternatives, due to erroneous acceptance errors like those we encountered, and some legitimate alternatives will be absent due to erroneous rejection errors. For both speed and accuracy we recommend the use of a dictionary solution.

Table 6. Examples (from our dictionary) that show alternative females names derived from a male name.

| Name | Alt1 | Alt2 | Alt3 | Alt4 | count | Standard Name |
|---|---|---|---|---|---|---|
| بدوي | بدوى | بديوي | | | 195 | بدوى |
| بدويه | أبدويه | بدوه | بديه | بديويه | 87 | بدويه |
| بشير | ابشير | البشير | | | 4045 | بشير |
| بشيرا | ابشيره | بشيره | | | 11 | بشيره |
| رزيق | ارزيق | رزق | | | 7 | رزق |
| رزيقه | ارزيقه | رزقه | | | 30 | رزقه |
| رميس | روميس | | | | 34 | رميس |
| رميساء | الرميساء | روميساء | | | 38 | رميساء |
| رهيف | رهف | | | | 17 | رهف |
| رهيفا | رهيف | رهيفاء | رهيفه | روهيفا | 15 | رهيفه |
| سحيم | اسحيم | | | | 4 | اسحيم |
| سحيما | سحمه | سحيمه | اسحيمه | | 5 | سحيمه |
| صبحي | صبحي | صبيحي | صبيحى | | 3 | صبحي |
| صبحه | اصبيحه | صبحه | صبيحا | صبيحى | 573 | صبحه |

Indexing the field of the name in the dictionary speeds up searching of the dictionary for alternatives. The cost is O(log k), where k is the number of names in the dictionary (17992 in our case). Searching the whole targeted database (which is normally indexed by name) costs O(m log n), where m is the number of alternatives, which varies from 1 to 13, and n is the number of names in the target database. Thus the total time complexity is equivalent to O(log k) + O(m log n). Because k and m are both constants, time complexity can be approximated to O(log n). This means we did not add significant time to normal search algorithms used in the target database.

Manually work on the dictionary took us more than 500 working hours; still we do not guarantee the absence of all errors. Therefore, we will publish the dictionary on the Internet for public use, so that it can be read, but also enhanced by users. Users can thus add new names and alternatives, or even delete mistakenly accepted alternatives. After user input, the dictionary will be ready for efficient use with different databases. Currently, the dictionary can be used within real-world databases with the caution that manual editing does not guarantee 100% correctness.

**5. Conclusion**

In this work, we introduce a new dictionary of Jordanian first names and their different writing forms to be used by system developers, database administrators, and researchers. This dictionary is meant to solve the problem of multiple alternative written forms of names in Arabic. The problem becomes more complicated when searching databases for such names. We created the dictionary by applying rules that govern why and how different forms of the same Arabic name are created.

Due to different types of errors and different reasons for problems, we needed to manually edit the dictionary to remove errors. During our review, we identified a number of reasons for errors, such as spelling errors and typos introduced by data-entry clerks, as well as inconsistent rendering of compound names. Therefore, we encourage database administrators to train data-entry clerks not only in how to use the system, but also, and more importantly, how to write correctly.

When a user searches a database for an Arabic name, the dictionary is invoked to provide all alternatives (if any) to be searched for in the database. This frees the user from needing to enter all alternatives (if the user is even aware of all alternatives) of the name. Our solution is also important for other problems, such as document matching, translation, name entity problem, and NLP in general. Moreover, the dictionary contains a standard written form for each name, which is the most common form of the name.

Because we edited the dictionary manually, we cannot claim that the solution is ready for use as-is; rather, it needs further editing, and more efforts by other researchers to include other names from different Arab countries before it can be considered a comprehensive solution ready for use with different database systems.

In future work, we intend to extend the dictionary to include names from other Arab countries as well as other languages that use Arabic text, such as Persian and Urdu, because many names in both languages are similar to those in Arabic. The methodology, with some different rules, might even be applied to other languages such as English, because a similar problem may occur, with a greater or lesser degree of complexity, for many languages.